\documentclass[10pt,aps,twocolumn]{revtex4-2}
\usepackage{bm}
\usepackage{amsmath}
\usepackage{graphicx}
\usepackage{hyperref}
\usepackage{mathtools}

\begin{document}

\title{Predicting Quadrupole deformation via anisotropic flow and transverse momentum spectra in isotopic  $\mathbf{\prescript{128-135}{54}{\mathrm{Xe}}}$ collisions at LHC}
\date{\vspace{-15ex}}

\author{Saraswati Pandey$^{1}$}
\email{saraswati.pandey13@bhu.ac.in}
\author{B. K. Singh$^{1,2}$}%
\email{bksingh@bhu.ac.in}
\affiliation{$^{1}$Department of Physics, Institute of Science, Banaras Hindu University (BHU),Varanasi, 221005, INDIA} 
\affiliation{$^{2}$Discipline of Natural Sciences, PDPM Indian Institute of Information Technology Design $\&$ Manufacturing, Jabalpur-482005, India} 

\begin{abstract}
\noindent
In the hydrodynamical description of heavy-ion collisions, the elliptic flow $\mathrm{v_{2}}$ and triangular flow $\mathrm{v_{3}}$ are sensitive to the quadrupole deformation $\mathrm{\beta_{2}}$ of the colliding nuclei. We produce $\mathrm{v_{2}}$ and $\mathrm{v_{3}}$ ratios qualitatively and quantitatively in most-central Xe-Xe collisions at 5.44 TeV. By employing HYDJET++ model, we study the sensitivity of anisotropic flow coefficients and mean transverse momentum to the quadrupole deformation and system-size in isotopic Xe-Xe collisions. Flow observables strongly depend on the strength of nucleon-nucleon scattering occuring in even-A and odd-A nuclei. Flow for odd-A nuclei is suppressed in comparison to flow in even-A collisions. There exists a linear inter-dependence between $\mathrm{p_{T}}$ integrated anisotropic flow and nuclear deformation. Mean transverse momentum signifies the fireball temperature in body-body and tip-tip collisions. There exists a negative linear correlation of $\mathrm{\langle p_{T} \rangle}$ with collision system-size and a positive correlation with nuclear deformation. Flow measurements in high-energy, heavy-ion collisions using isotopic collision systems, offer a new precision tool to study nuclear structure physics. Observation of nuclear structure properties like nuclear deformation in a heavy-ion collision such as this would be very interesting.
\end{abstract}

\date{\today}

\maketitle 

\section{Introduction}
\label{intro}

Atomic nuclei are characterized by several intrinsic properties, one of them being deformed shapes arising as a result of high precision hyperfine studies in the field of atomic physics, thereby marking a non-spherical nuclear charge distribution \cite{PhysRevLett.128.022301, RevModPhys.83.1467, Heyde_2016}. Such atomic nuclei in their ground state are deformed from a well defined spherical shape. The deformation has a significant dependence on the number of protons and neutrons, particularly in the proximity of a full shell or subshell, reflecting collective motion induced by interaction between valence nucleons and shell structure \cite{PhysRevC.105.014905}. This nuclear deformation emerges from the short-range strong nuclear interaction between the nucleons \cite{RevModPhys.83.1467, Heyde_2016, PhysRevC.105.044905, PhysRevLett.117.172502}. This lead to a plethora of research on the topics such as shape evolution, quadrupole deformation, triaxiality or shape coexistence, octupole deformation, hexadecapole deformation, etc \cite{RevModPhys.83.1467, PhysRevLett.117.172502, Frauendorf_2018, Zhou_2016}.

In recent decades, Relativistic Heavy Ion Collider (RHIC) at the BNL and Large Hadron Collider (LHC) at the CERN have conducted collider experiments, developing an understanding that the deformation of the colliding nuclei strongly affect the QGP observables studied in (ultra-) relativistic heavy-ion collisions \cite{Singh:2017fgm, PhysRevC.103.014903, refId0}. One of the main motivations in this context has been elliptic flow $\mathrm{v_{2}}$ \cite{doi:10.1146/annurev-nucl-102212-170540}. In (ultra-)relativistic nucleus-nucleus collisions, elliptic flow arises as a consequence of the quadrupole asymmetry (ellipticity) of the system created upon collision, in the plane transverse to the beam direction \cite{PhysRevC.83.064904}. When two nucleus collide at a certain impact parameter, the overlap area has an elliptical deformation and we observe that $\mathrm{v_{2}}$ grows sharply with the collision impact parameter. However, for most of the isotopes, even in the case of vanishing impact parameter, it is expected that $\mathrm{\epsilon_{2}\neq 0}$, and therefore $\mathrm{v_{2}\neq 0}$ from nuclear structure domain becomes argumentative as most of the nuclei exhibit a non-vanishing intrinsic quadrupole moment, or to say, an ellipsoidal deformation \cite{PhysRevC.104.L041903}.

In ultra-relativistic heavy-ion collisions, the dynamical response of the QGP to its initial spatial anisotropy produces anisotropic flow. This initial spatial anisotropy is affected by the geometric shape of the colliding nuclei, thereby, forming an intrinsic relationship between the structure of atomic nuclei and the phenomenology of heavy-ion collisions. Thus, if we match high-energy data to the data at low-energy expectations, we can assess whether our knowledge of nuclear physics across energy scales yields consistent results. Anisotropic flow ($\mathrm{v_{n}}$) plays a major role in elucidating the final-state effect as it constrains the viscous hydrodynamic response to the eccentricity ($\mathrm{\epsilon_{n}}$) of the energy-density distribution produced in the initial stages of the collision \cite{PhysRevLett.86.402, PhysRevLett.89.212301, PhysRevC.66.054905, HEINZ2002269, HIRANO2006299, PhysRevLett.99.172301, PhysRevC.81.034909, PhysRevC.93.064901, PhysRevC.86.044908, PhysRevLett.109.202302, Lacey_2016}. The initial-state profile for each of the colliding nuclei is characterized by the woods-Saxon distribution for the nuclear density as:

\begin{equation*}
\mathrm{\rho(r,\theta,\phi) = \frac{\rho_{0}}{1 + \exp ([r-R^{'}(\theta,\phi)]/a)}}, 
\end{equation*}
\begin{equation}
\mathrm{R'(\theta,\phi) = R[1 + \beta_{2} Y_{2}^{0} + \beta_{3} Y_{3}^{0} + \beta_{4} Y_{4}^{0} + ...]},
\label{eq1}
\end{equation} 

where, $\rm{R=R_{0} A^{1/3}}$ is the nuclear radius ($\rm{R_{0}} = 1.15$fm), $\mathrm{\rho_{0}}$ is the center of the nucleus density, a is the surface thickness (i.e., the effective diffusivity), R$'$ is the nuclear radius, $Y_{n}^{0}$ are the spherical harmonics corresponding to each harmonic order, and $\mathrm{R'(\theta,\phi)}$ is the nuclear surface with relevant axial symmetric quadrupole ($\mathrm{\beta_{2}}$), octupole ($\mathrm{\beta_{3}}$) and hexadecapole ($\mathrm{\beta_{4}}$) deformations \cite{PhysRevC.105.014905, PhysRevLett.127.242301, PhysRevC.87.044908, PhysRevC.106.L031901, PhysRevLett.128.022301, PhysRevC.105.014906}. In this study, we have ignored all higher moment orders of nuclear deformation and only focussed on quadrupole moment. The other parameters such as $\rm{R}$ and $\rm{a}$ are fixed in case of every nuclei. Positive $\mathrm{\beta_{2}}$ describes the overall quadrupole deformation. Nuclear deformation is a fundamental property of the atomic nucleus. It explains the correlated behaviour of the associated dynamics of nucleons. Many atomic nuclei have quadrupole or octupole deformation, which may influence the anisotropic flow coefficients$\mathrm{'}$ magnitude, its fluctuations, and correlations. Recent measurements presented detailed comparisons between Au+Au and U+U collisions \cite{PhysRevLett.115.222301} as well as Pb+Pb and Xe+Xe collisions \cite{elliptic_2022, 201882}. The studied observables were found compatible with nuclear deformation. However, the analysis performed so far on flow harmonics is not so successful in providing constraints useful for the detailed characterization of the deformation in the colliding nuclei.

An important aspect of the above study is that the overprediction or underestimation of experimental anisotropic flow data from theoretical calculations in larger collision systems signifies the missing physics. For instance, significant enhancement was observed in $\mathrm{v_{2}}$ \{2\} in deformed central $\mathrm{^{129}}$Xe-$\mathrm{^{129}}$Xe collisions compared to that in case of spherical $\mathrm{^{129}}$Xe nucleus. $\mathrm{^{129}}$Xe nucleus is proposed to have a small quadrupole deformation, and the enhancement in flow can be seen in the articles \cite{PhysRevC.103.014903, PhysRevC.97.034904, 201882, 2019166, Pandey_2022}. In another study, $\mathrm{^{238}}$U-$\mathrm{^{238}}$U collisions with the nuclei having a substantial quadrupole deformation, also were observed to have an increase in the elliptic flow in ultra-central events \cite{PhysRevLett.115.222301, PhysRevC.92.044903, PhysRevC.95.064907, PhysRevC.99.024910}. Thus, flow studies are heplful in mapping out the quadrupole deformations evoluting in the series of stable $\mathrm{_{54}}$Xe isotopes. Now, of course, studying xenon isotopic chain over another nuclei such as samarium isotopes can be argumentative. The reason is atrributed to the fact that, firstly, in 2017, LHC carried out an eight hour run of deformed Xe-Xe collisions at 5.44 TeV. The mass number of Xe nucleus lies in mid-between p and Pb$^{208}$. Thus, collisions of Xe$^{129}$ nuclei would bridge the multiplicity gap between the larger Pb-ion systems and smaller systems like p+p and p+Pb. Back then, Xe was considered somewhat prolate. This deformation in xenon allows us to probe a different initial condition. However, further research confirmed Xe$^{129}$ to be triaxial. So, an odd isotope of intermediate-mass nuclei has already made a significant research impact in the field of heavy-ion physics. Secondly, Xe$^{129}$ is triaxial while Xe$^{128}$ is prolate. Differing in shape, xenon with A (mass) changing to even and odd, it becomes obligatory to study a series of even and odd isotopes and be investigated in heavy-ion collisions. Also, $\beta_{2}$ values for odd- and even-A xenon nuclei showed a lot of variation in low energy structure models. Thus, one is motivated to cover a range of both the kinds of nuclei with some significant value of quadrupole moment. Thirdly, in case of a recent study on samarium (Sm) isotopes, it was seen that as the mass number increased, the quadrupole moment increased \cite{bally2022imaging}. However, in case of xenon, the situation was quite the opposite for the range of isotopes chosen for study, thereby making this set of isotopic chain important to be studied. Fourthly, a huge amount of research is going on to study isobaric collisions at RHIC and LHC energy regime where Ru-Ru and Zr-Zr collisions from STAR experiment have become quite interesting area of research. A similar study to perform at LHC energies could be with Nd-Sm pair as suggested in article \cite{bally2022imaging}. However, these are even-A nuclei. But, odd-A Xe$^{129}$ has a strong background of research in heavy-ion collisions. So, why not pick the odd-A, Xe-Sm pair. We preserve this for future work, but as the full isotopic chain of eight isotopes of Sm have been studied in heavy-ion collisions, it makes Xe worthy enough to be also studied through heavy-ion collisions. The hydrodynamic response is essentially expected to be unchanged over the isotopic chain, however, these collision systems offer a strong lever-arm to probe in detail how the initial condition of QGP responds to varying nuclear shapes. Such a study in high-energy heavy-ion collisions would give new experimental insights onto the role and effect of quadrupole deformations in such nuclei.

Most-central collisions offer the best scenario to test the sensitivity of nuclear deformation \cite{PhysRevLett.124.202301, PhysRevC.102.024901}. This is because at zero impact parameter the anisotropy happens not only due to fluctuations but also due to the non-spherical shape, while in other collision centralities the impact region is dominated by the almond shape of the overlap region \cite{PhysRevC.102.054905}. However, in most-central collisions, the shape contribution is quite smaller in comparison to the contribution from fluctuations. Moreover, other possibilities causing fluctuations, which scale with the collision system size, affect minimally in most-central collisions where the overlap region upon collision is largest. Also, the large amount of energy deposited in these most-central collisions form a hot and dense quark-gluon plasma (QGP) \cite{doi:10.1146/annurev-nucl-101917-020852} in the overlap region, whose shape and size are strongly correlated with nuclear deformation.

In this paper, we aim to address the question of whether the values of $\mathrm{\beta}$ found in low-energy literature are consistent with $\mathrm{v_{2}}$ and $\mathrm{v_{3}}$ data at high energy. We investigate this using a simple Monte Carlo HYDJET++ model. Further, we also aim to demonstrate that flow measurements in ultra-relativistic heavy-ion collisions using isotopic Xe systems, offer a new precision tool to study nuclear structure physics. We apply the idea to run isotopes $\mathrm{\prescript{128-135}{54}{\mathrm{Xe}}}$, colliding symmetrically and make predictions on ratios of $\mathrm{v_{n}}$ from the known values of quadrupole deformation $\mathrm{\beta_{2}}$ from nuclear structure measurements, as given in table \ref{table}. Here, we have ignored higher harmonic orders of nuclear deformations. Further, we will also study the transverse momentum as a function of the nuclear deformation. We also aim to study the observables in two extreme geometries of deformed collision systems. In the next section \ref{model}, we discuss the essentials of Monte Carlo HYDJET++ model followed by anisotropic flow presented in the section \ref{flow}. Further, we will present our results and discussions in section \ref{results} followed by the summary of our work in section \ref{summary}.

\section{HYDJET++ Model}
\label{model}

To understand the hydrodynamic response to nuclear deformations and make predictions, we employ a Monte Carlo HYDJET++ (hydrodynamics plus jets) event generator \cite{LOKHTIN2009779} developed to study heavy-ion collisions at RHIC as well as LHC energies. It performs simulation by superimposing soft (hydro-type) state and the hard state (resulting from multi-parton fragmentation) simultaneously, treating both independently. It meticulously treats soft hadroproduction (collective flow phenomenon and the resonance decays) as well as hard parton production, also examining the known medium effects (jet quenching and nuclear shadowing). The in-depth details of the model and the procedure of simulation can be found in the corresponding articles \cite{LOKHTIN2009779, lokhtin2009hydjet++} and the references there within. The model parameters have been tuned to reproduce the experimental LHC data on various physical observables measured in Xe-Xe collisions at 5.44 TeV of center-of-mass energy per nucleon. In brief, the model is described as follows:

\begin{table}[h]
\begin{tabular}{|r|c|} \hline
\quad Nuclei \quad \quad & $\mathrm{\beta_{2}}$ \\ \hline
$\prescript{128}{54}{\mathrm{Xe}}$ \quad \quad & \quad \quad 0.1837 \quad \quad \\ \hline

$\prescript{129}{54}{\mathrm{Xe}}$ \quad \quad & \quad \quad 0.1620 \quad \quad \\ \hline

$\prescript{130}{54}{\mathrm{Xe}}$ \quad \quad & \quad \quad 0.1690 \quad \quad \\ \hline

$\prescript{131}{54}{\mathrm{Xe}}$ \quad \quad & \quad \quad 0.1400 \quad \quad \\ \hline

$\prescript{132}{54}{\mathrm{Xe}}$ \quad \quad & \quad \quad 0.1409 \quad \quad \\ \hline

$\prescript{133}{54}{\mathrm{Xe}}$ \quad \quad & \quad \quad 0.1220 \quad \quad \\ \hline

$\prescript{134}{54}{\mathrm{Xe}}$ \quad \quad & \quad \quad 0.1200 \quad \quad \\ \hline

$\prescript{135}{54}{\mathrm{Xe}}$ \quad \quad & \quad \quad 0.0830 \quad \quad \\ \hline
\end{tabular}
\caption{Values of $\mathrm{\beta_{2}}$ deformations are calculated using reduced transition probabilities \cite{2001RA27}, finite-range droplet macroscopic (FRDM) and the folded-Yukawa single-particle microscopic nuclear-structure models mentioned in article \cite{MOLLER20161}, and some calculated quadrupole deformation parameters from \cite{2005ST24, 2005IS09}.}
\label{table}
\end{table}

The hard state of an event in HYDJET++ is treated using Pythia Quenched (PYQUEN) model \cite{lokhtin2006model}. PYQUEN model modifies a jet event produced by PYTHIA by producing nucleonic collision vertices in accordance with the Glauber model at a certain impact parameter. This is followed by a rescattering-by-rescattering simulation of the parton path in the dense zone and the associated radiative and collision energy losses \cite{PhysRevD.27.140, PhysRevD.44.R2625, Lokhtin:2000wm, PhysRevC.60.064902, PhysRevC.64.057902}. Then the final hadronization is carried out according to the Lund String Model \cite{andersson2005lund} for hard partons and in medium emitted gluons. An impact parameter-dependent parameterization obtained under the framework of Glauber-Gribov theory is used to incorporate the medium effect called nuclear shadowing \cite{TYWONIUK2007170, gribov1969glauber}. The hard state in HYDJET++ is separated from the soft state by a free parameter named  $\mathrm{p_{T}^{min}}$. Events for which generated total transverse momentum is greater than $\mathrm{p_{T}^{min}}$ are considered as the hard part whereas events for which  $\mathrm{p_{T}^{min}}$ are considered as the soft part.

The soft state of an event in HYDJET++ is a thermal hadronic state generated on the chemical and thermal freeze-out hypersurfaces obtained from a parameterization of relativistic hydrodynamics with preset freeze-out conditions \cite{PhysRevC.74.064901, PhysRevC.77.014903}. Here, it is assumed that the hadronic matter produced in a nuclear collision reaches a local equilibrium after a short period of time (\textless 1fm/c) and then expands hydrodynamically. Basically, the hadronic matter created in heavy-ion collisions is a hydrodynamically expanding fireball with the equation of state of an ideal hadron gas. The total yield of particular particle species is determined by the freeze-out temperature, chemical potential and by the total co-moving volume or effective volume of particle production which is a functional of the field of collective velocities on the hypersurface. The effective volume utilizes the collective velocity profile and the form of hypersurface canceling out in all particle number ratios. This effective volume can be used to calculate the hadronic composition at both chemical and thermal freeze-outs \cite{AKKELIN2002439}. At the former one, which happens soon after hadronization, the chemically equilibrated hadronic composition is assumed to be established and frozen in further evolution. This implies that the chemical concentrations of hadron gas do not change during the evolution. This is justified as the rate of expansion in hadron gas is larger than the rate of inelastic reactions and lesser than that of elastic collisions \cite{muller1986, Rafelski1996}. In HYDJET++, the presumption of single freeze-out is eliminated as the particle densities at the chemical freeze-out stage are too high to consider the particles as free streaming \cite{AKKELIN2002439}. Therefore, a more complex scenario of different thermal and chemical freeze-outs(T$\mathrm{_{ch} \geq}$ T$\mathrm{_{th}}$) is opted. The system expands hydrodynamically with frozen chemical composition in between these two freeze-outs, then cools down and the hadrons stream freely as soon as the thermal freeze-out temperature is reached. In non-central collisions, the shape of the emission region in the transverse (x-y) plane is approximately elliptical with the (z-x) plane coinciding with the reaction plane. The two radii of the ellipse are: $R_{x}(b)=R_{s}(b)\sqrt{1-\epsilon(b)}$ and $R_{y}(b)=R_{s}(b)\sqrt{1+\epsilon(b)}$, where $\epsilon(b)=(R^{2}_{y}-R^{2}_{x})/(R^{2}_{x}+R^{2}_{y})$ and the scale factor $R_{s}=\sqrt{(R^{2}_{x}+R^{2}_{y})/2}$. Then, the transverse radius $R(b,\phi)$ of the fireball on the azimuthal angle is:
\begin{equation}
R(b,\phi)=R_{s}(b)\frac{\sqrt{1-\epsilon^{2}(b)}}{1+\epsilon(b)\cos 2\phi},
\end{equation}   
where,
\begin{equation}
R_{s}(b)=R_{s}(b=0)\sqrt{1-\epsilon_{s}(b)}.
\end{equation}
This means that the dimensionless ratio $R_{s}(b)/R_{s}(0)$ at the freeze-out moment depends on the collision energy, the radius $R_{A}$ of the colliding (identical) nuclei and the impact parameter b through a dimensionless $\epsilon_{s}(b)$ only.

\subsection{Anisotropic flow in HYDJET++}
\label{flow}

Quark-Gluon Plasma, the deconfined state of color charges is believed to be created in relativistic heavy-ion collisions \cite{singh2000quark}. The pressure gradients generated in the QGP medium via most probably non-central collisions at high relativistic energies convert the initial anisotropies to momentum anisotropies of the produced particles via multiple interactions. This phenomenon is called anisotropic flow. This anisotropic flow is defined by the coefficients from the fourier expansion of the azimuthal distribution of the produced particles \cite{voloshin1996flow, poskanzer1998methods} as-

\begin{equation}
\mathrm{\dfrac{dN}{d\phi} \propto 1+ 2\sum\limits_{n=1}^{\infty} v_{n}\cos[n(\psi-\psi_{R})]}
\end{equation}
$\text{where}$,\\

$\mathrm{\psi} \text{= azimuthal angle of the produced particle}$,\\

$\text{n = harmonic value, and}$\\

$\mathrm{\psi_{R} = \text{reaction plane.}}$ \\

For n=2, we have the second order coefficient called as elliptic flow, which is sensitive to the early evolution of the system. In HYDJET++ model, the reaction plane for even order flow is zero for each event. The elliptic flow coefficient $\rm{v_{2}}$ is determined in the hadron distribution over the azimuthal angle $\rm{\psi}$ relative to the reaction plane $\rm{\psi_{R}}$, such that,

\begin{equation}
\mathrm{v_{2} = \langle \cos[2(\psi-\psi_{R})] \rangle}, \text{with } \mathrm{\psi = \tan^{-1} ({p_{y}/p_{x}})}.
\end{equation} 

The above equation in the simplified form for $\rm{v_{2}}$ in terms of particle momenta is expressed by-

\begin{equation}
\mathrm{v_{2} = \left \langle \frac{p_{x}^{2}-p_{y}^{2}}{p_{x}^{2}+p_{y}^{2}} \right \rangle = \left \langle\frac{p_{x}^{2}-p_{y}^{2}}{p_{T}^{2}} \right \rangle} \quad
\end{equation}

The soft particle emission which contributes most to the elliptic flow, arises from the freeze-out hypersurface. To simulate higher azimuthal anisotropy harmonics, parameterization of the fluid velocity corresponding to each harmonic on the freeze-out surface is performed in HYDJET++. The model does not describe the evolution of the fireball from the initial to the final freeze-out stage. It utilizes simple parameterization of the freeze-out hypersurface. The anisotropic elliptic shape of the initial overlap of the colliding nuclei results in a corresponding anisotropy of the outgoing momentum distribution. The second harmonic $\rm{v_{2}}$ is described using the coefficients $\rm{\epsilon_{2}}$(b) and $\rm{\delta_{2}}$(b). These are the second order spatial anisotropy and momentum anisotropy parameters, respectively. $\rm{\epsilon_{2}}$(b) exemplifies the elliptic modulation of the final freeze-out hypersurface at a given impact parameter b, whereas $\rm{\delta_{2}}$(b) deals with the alteration of flow velocity profile. These two parameters can be treated independently for each centrality or can be made interdependent via the dependence on the initial ellipticity $\rm{\epsilon_{0}(b)=b/2R_{A}}$ where $\rm{R_{A}}$ is the nucleus radius. Here, we are treating them independent of each other.

The spatial anisotropy gets transformed into momentum anisotropy at freeze-out, because each of the fluid cells is carrying some momentum. In case where $\rm{\delta_{2}(b) \neq 0}$ even the spherically symmetric source can mirror the spatially contracted one. The elliptic flow coefficient $\rm{v_{2}(\epsilon_{2}, \delta_{2})}$ in the hydrodynamical approach (incorporated in HYDJET++ framework) \cite{LOKHTIN2009779} is given as:
 
\begin{equation}
\rm{v_{2}(\epsilon_{2},\delta_{2}) \propto \frac{2(\delta_{2}-\epsilon_{2})}{(1-\delta_{2}^{2})(1-\epsilon_{2}^{2})}}.
\end{equation}  

For triangular flow $\rm{v_{3}}$ in HYDJET++, the model has another parameter $\rm{\epsilon_{3}}$(b), for spatial triangularity of the fireball. Thus, the modified radius of the freeze-out hypersurface in azimuthal plane reads:

\begin{equation}
\rm{R(b,\phi)=R_{ell}(b)\lbrace1+\epsilon_{3}(b)\cos[3(\phi-\psi_{3}^{RP})]+...\rbrace}.
\end{equation}           
where,
$\rm{\phi}$ = spatial azimuthal angle of the fluid element relative to the direction of the impact parameter.

The phase $\rm{\psi_{3}^{RP}}$ gives us the advantage to introduce a third harmonic having its own reaction plane, distributed randomly with respect to the direction of the impact parameter ($\rm{\psi_{2}^{RP}=0}$). Such modifications do not affect the elliptic flow (controlled by $\rm{\epsilon_{2}(b)}$ and $\rm{\delta_{2}(b)}$). Hence, the triangular dynamical anisotropy can be incorporated by the parameterization of the maximal transverse flow rapidity \cite{refId02006},

\begin{equation}
\rm{\rho_{u}^{max}(b)=\rho_{u}^{max}(0)\lbrace 1+ \rho_{3u}(b)\cos [3(\phi-\psi_{3}^{RP})] +...\rbrace}.
\end{equation}  

Hence, we can calculate higher harmonics with respect to the direction of the impact parameter b.

\begin{figure}[htbp]
\includegraphics[width=8.0cm,height=6.2cm, keepaspectratio]{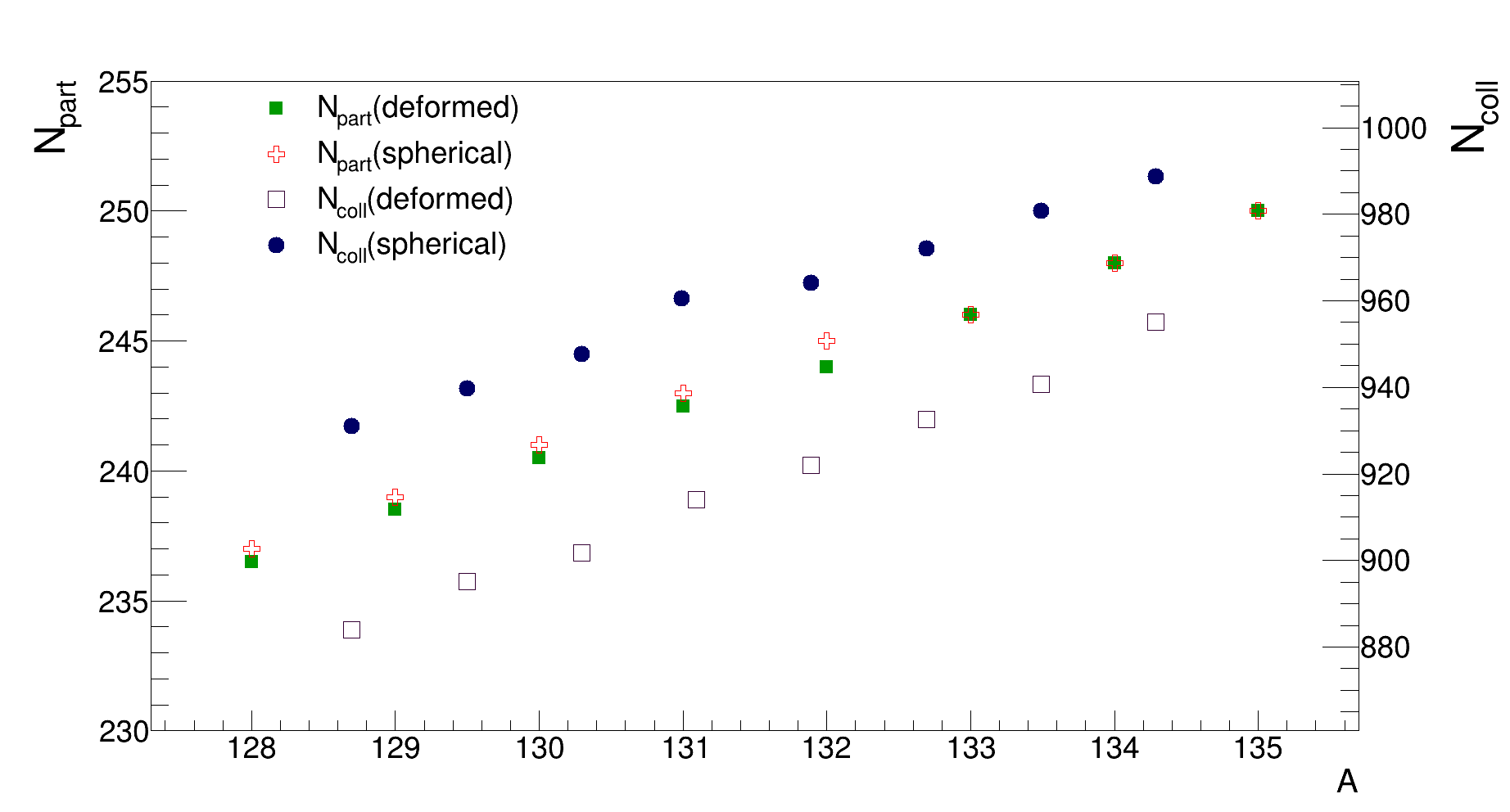} 
\caption{Variation of number of participants $\rm{N_{part}}$ and number of binary collisions $\rm{N_{coll}}$ with atomic weight  (A) in most-central (0-5)\% Xe-Xe collisions at various values of nuclear deformation parameter ($\rm{\beta_{2}}$).}
\label{npartncoll}
\end{figure}

HYDJET++ model provides reasonable description of anisotropic flow in heavy-ion (small, intermediate, and large) collision systems at both RHIC and LHC energies \cite{Singh:2017fgm, PhysRevC.103.014903, refId0}. The results show proper qualitative as well as quantitative explanation for the flow in spherical and deformed collision systems. We simulate $\mathrm{10^{5}}$ minimum bias, symmetric, and isotopic (0–5)\% most-central $\prescript{128-135}{54}{\mathrm{Xe}}$ collisions under HYDJET++ framework with $\mathrm{R_{0}=1.15A^{1/3}}$fm and surface thickness (i.e., the effective diffusivity) $\mathrm{a=0.59 \pm 0.07}$fm. The observables are calculated under $\mathrm{ \vert \eta \vert < 0.8}$ and $\mathrm{0<p_{T}<5.0}$ $\mathrm{GeV/c}$ kinematic ranges. Under these circumstances, the number of participants $\mathrm{N_{part}}$ and the number of binary collisions $\mathrm{N_{coll}}$ as a function of atomic mass have been shown in figure \ref{npartncoll}, thereby certifying our model. It is observed that $\mathrm{N_{part}}$ and $\mathrm{N_{coll}}$ linearly increase with atomic mass. This is because, HYDJET++ model uses Glauber-type initial condition for soft (hydro-type) part. This atomic mass, if we see table \ref{table}, is almost inversely proportional to nuclear deformation. Spherical collisions have higher $\mathrm{N_{part}}$ and $\mathrm{N_{coll}}$ compared to deformed collision systems. The difference in the number of participants of deformed and spherical isotopic collisions is significantly visible at lower atomic mass but eventually decreases as one moves to higher A. The model parameters which control the execution of our Monte Carlo generator show suitable match with experimental data at LHC as presented in the recent articles \cite{PhysRevC.103.014903, refId0}. The model input parameters for $\mathrm{^{129}}$Xe-$\mathrm{^{129}}$Xe collisions at 5.44 TeV are provided in table \ref{table2}.

\begin{table}[ht]
\centering
\begin{tabular}{|r|c|} \hline
\quad Input Parameter \quad \quad & Value \\ \hline
$\mathrm{T_{ch}}$ \quad \quad & \quad \quad 151 MeV \quad \quad \\ \hline

$\mathrm{T_{th}}$ \quad \quad & \quad \quad 105 MeV \quad \quad \\ \hline

$\mathrm{\mu_{th}}$ \quad \quad & \quad \quad 0 \quad \quad \\ \hline

$\mathrm{\mu_{B}}$ \quad \quad & \quad \quad 0 \quad \quad \\ \hline

$\mathrm{\mu_{s}}$ \quad \quad & \quad \quad 0 \quad \quad \\ \hline

$\mathrm{p_{T}^{min}}$ \quad \quad & \quad \quad 10.5 GeV/c \quad \quad \\ \hline
\end{tabular}
\caption{Model parameters for $\mathrm{^{129}}$Xe-$\mathrm{^{129}}$Xe collisions at 5.44 TeV in most central class of collisions.}
\label{table2}
\end{table} 
These input model parameters which control the execution of the Monte Carlo generator are obtained by simulating (0–5)\% most-central $\mathrm{^{129}}$Xe-$\mathrm{^{129}}$Xe collisions and then matching it with the experimental data at LHC.

\begin{figure}[htbp]
\includegraphics[width=8.0cm,height=6.5cm]{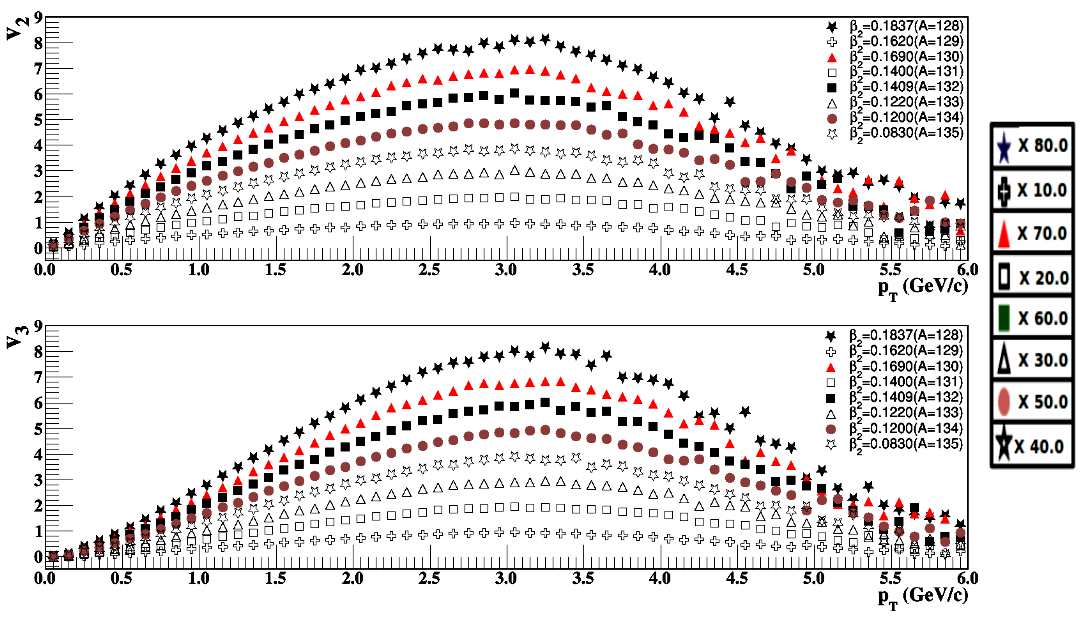}
\caption{Transverse momentum dependence of anisotropic flow $\mathrm{v_{n}}$ (n=2,3) in deformed Xe-Xe collisions at 5.44 TeV center of mass energy using HYDJET++ model. The figure shows flow for eight isotopes (A=128-135) of Xenon with different values of quadrupole moment.}
\label{v2v3flow}
\end{figure}

\section{Results and Discussion}
\label{results}

The upper panel of figure \ref{v2v3flow} shows elliptic flow as a function of transverse momentum ($\mathrm{p_{T}}$) for different isotopic minimum bias Xe-Xe collisions with different prolate-type deformations. The lower panel presents triangular flow for the same scenario. Here, the bin values have been scaled by some factor shown in the figure, to visualize the difference between the flow of different isotopic collisions. Using the same factors, the scaling in the other results also have been introduced.

It is observed that, for even isotopes, anisotropic flow $\mathrm{v_{n}}$ (n=2,3) decreases as atomic mass increases and quadrupole deformation decreases. However, for odd isotopes, situation appears to be quite opposite; anisotropic flow increases as atomic mass increases and quadrupole deformation decreases. Thus, for even isotopes, anisotropic flow is directly proportional to nuclear deformation whereas for odd isotopes, this flow is inversely proportional to nuclear quadrupole moment. Thus, for an odd-A isotopic chain, the relationship between nuclear deformation and anisotropic flow (as in magnitude) is opposite to that of the even-A isotopic chain. This hints us at the strength of nucleon-nucleon scattering taking place in the two types of isotopic Xe-Xe collisions. The flow for odd-A nuclei is suppressed compared to even-A nuclei. Thus, $\mathrm{v_{n}}$-$\mathrm{p_{T}}$ correlation here directly connects us to the initial collision geometry and thereby back to the shape of the colliding nuclei. Elliptic flow $v_{2}$ is more than triangular flow, the difference being very less. Flow increases as we move from low $\mathrm{p_{T}}$ to high $\mathrm{p_{T}}$ region. The peak value at which flow is maximum is around $3.0$ GeV/c for elliptic flow. It shifts towards higher $\mathrm{p_{T}}$ as the harmonic value, n increases. 

\begin{figure}[htbp]
\includegraphics[width=9.0cm,height=6.5cm]{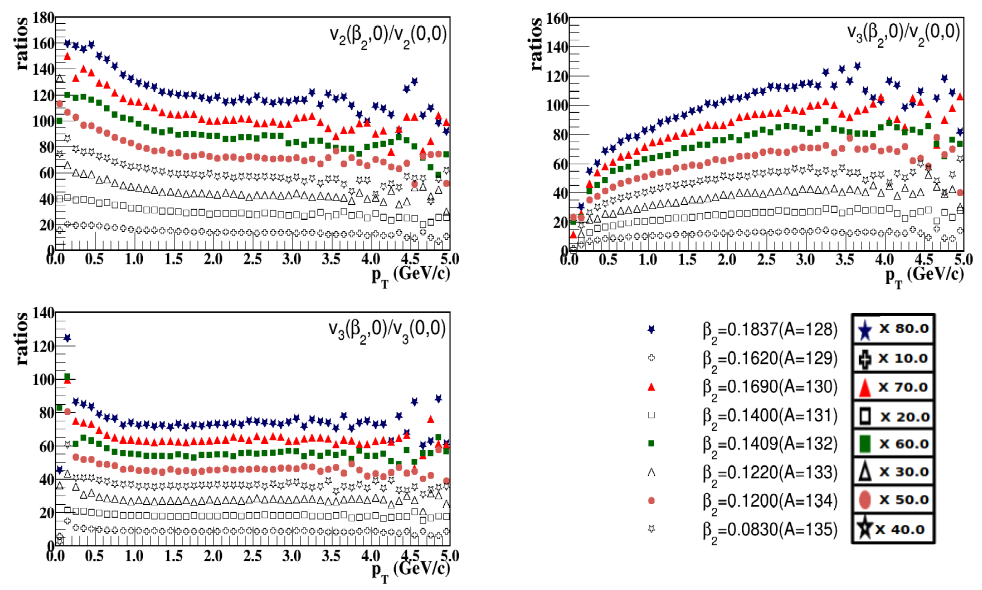}
\caption{Transverse momentum dependence of the ratios of anisotropic flow $\mathrm{v_{n}}$ (n=2,3) from deformed Xe-Xe collisions to their respective spherical Xe-Xe collisions at 5.44 TeV center of mass energy using HYDJET++ model. The figure shows flow for eight isotopes (A=128-135) of Xenon with different values of quadrupole moment.}
\label{flow_ratio}
\end{figure}

In figure \ref{flow_ratio}, we present ratios of $\mathrm{v_{n}(\beta_{2},\beta_{3}, ...)}$ for given values of $\mathrm{\beta_{2}}$ (with rest orders as zero) to that for spherical nuclei. In this scenario, three types of ratios are studied namely: (i) $\mathrm{v_{2}(\beta_{2},0)/v_{2}(0,0)}$, (ii) $\mathrm{v_{3}(\beta_{2},0)/v_{2}(0,0)}$, and (iii) $\mathrm{v_{3}(\beta_{2},0)/v_{3}(0,0)}$. Here, we observe that the ratios of elliptic flow and triangular flow  of deformed collisions with respect to their respective spherical collisions, show similar qualitative behaviour whereas the ratio of deformed $\mathrm{v_{3}}$ to spherical $\mathrm{v_{2}}$ has a different inference. Ratios (i) and (iii) present a somewhat linear relationship with increasing transverse momentum and gradually decrease towards higher $\mathrm{p_{T}}$ values. Higher quadrupole moment gives rise to higher flow for even-A isotopic collisions, hence, higher ratios. In odd-A isotopic collisions, the situation is reversed, higher deformation results in lower anisotropic flow, eventually giving smaller ratios. On the other hand, ratio (ii) of deformed $\mathrm{v_{3}}$ to spherical $\mathrm{v_{2}}$ presents a gaussian-type of behaviour. It increases at low $\mathrm{p_{T}}$ and reaches its zenith at around 3.0 GeV/c, then decreases (very slowly) towards higher transverse momentum. Thus, the figures bring into picture the effect of deformation and collision system-size for the various isotopic Xe-Xe collisions. The second ratio $\mathrm{v_{3} (\beta_{2} , 0)/v_{2} (0, 0)}$ presents a significant viscosity dependence. Viscous damping is smaller in $v_{2}$ than in $v_{3}$. Therefore, we observe that the ratio increases. Such an inspection with nuclear quadrupole moment suggests that for even-A nuclei, viscous damping increases with increasing nuclear quadrupole moment; while for odd-A nuclei, viscous damping decreases as nuclear deformation increases. Another important aspect is that the ratio is an important parameter for the $v_{2}-v_{3}$ puzzle. So, it is necessary to measure it as a function of nuclear deformation.

\begin{figure}[htbp]
\includegraphics[width=9.0cm,height=7.0cm]{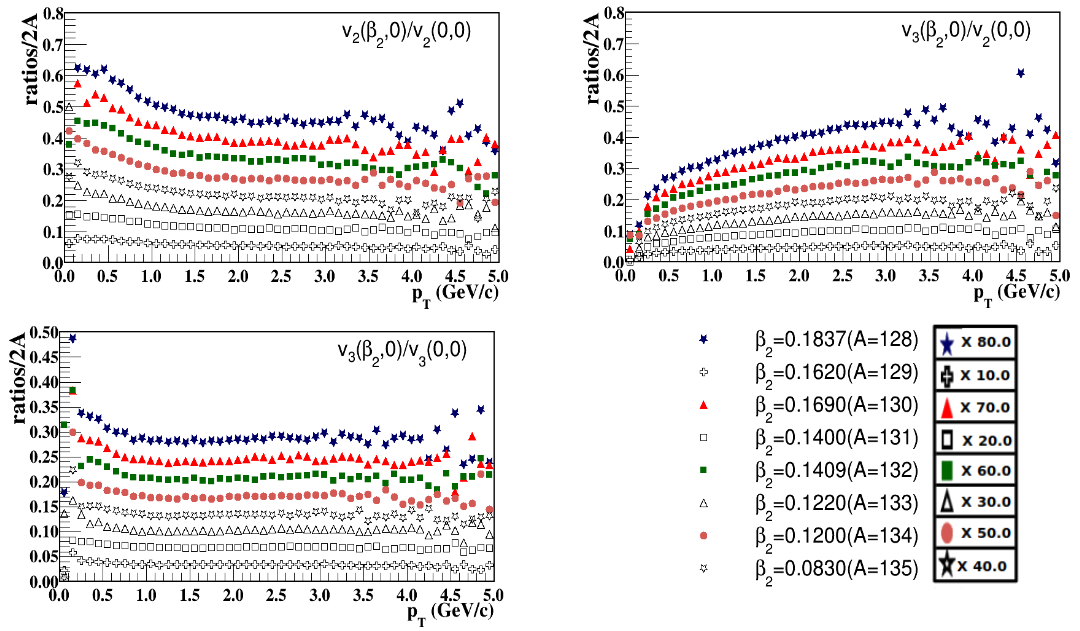}
\caption{Scaling of the ratios of anisotropic flow $\mathrm{v_{n}}$ (n=2,3) from deformed Xe-Xe collisions to their respective spherical Xe-Xe collisions at 5.44 TeV center of mass energy using HYDJET++ model by 2A. The figure shows flow for eight isotopes (A=128-135) of Xenon with different values of quadrupole moment.}
\label{ratio_scaling}
\end{figure}

In order to visualize the effect of nuclear deformation separately, we normalize our results with the total number of nucleon-pairs in each collision system (please see figure \ref{ratio_scaling}) and find the quantitative role of deformation in the three kinds of isotopic Xe-Xe collision systems, where we can easily ascertain the above inferences. Elliptic flow is larger than triangular flow is very evident from the above figures. The quantitative effect of quadrupole deformation irrespective of the atomic mass show the role of quadrupole moment on the initial nuclear density profile of nucleus, and hence on the initial geometry of the collision overlap region.

\begin{figure}[htbp]
\includegraphics[width=8.5cm,height=6.5cm]{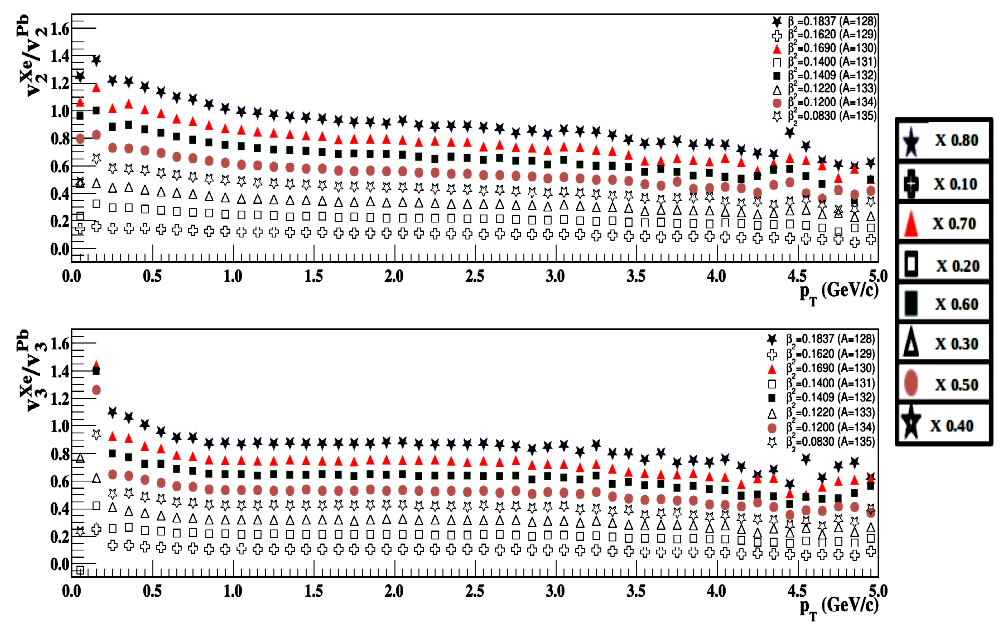}
\caption{Scaling prediction of anisotropic flow $\mathrm{v_{n}}$ (n=2,3) from deformed $\prescript{128-135}{54}{\mathrm{Xe}}$-$\prescript{128-135}{54}{\mathrm{Xe}}$ collisions to spherical $^{208}\mathrm{Pb}$-$^{208}\mathrm{Pb}$ collisions at 5.02 TeV center of mass energy using HYDJET++ model by 2A. The figure shows results for isotopes (A=128-135) of Xenon with different values of quadrupole moment.}
\label{Xe_ratio_Pb}
\end{figure}

To study the system-size dependence of anisotropic flow, figure \ref{Xe_ratio_Pb}, presents the ratio of flow observables in (deformed-) Xe-Xe collisions at 5.44 TeV to spherical Pb-Pb collisions at 5.02 TeV. A baseline expectation is that the flow observables should be equal in the larger (Pb) and smaller (Xe) collision systems. However, several effects break this scaling, and can be taken as corrections to the scale-invariant baseline. Here, we present this test for the various isotopic collisions as a function of transverse momentum $\mathrm{p_{T}}$. We observe that the flow strongly increases in central collisions for the Xe system  compared to the Pb system, which distinctly indicates about the nuclear deformation in Xenon. Based upon the strength of nuclear deformation and isotopic collision system-size, the ratio of each flow harmonic corresponding to each isotope exists quantitatively. The scaling shows that the effect of nuclear deformation is visible maximum in low $\mathrm{p_{T}}$ region and decreases as we move towards higher $\mathrm{p_{T}}$ values. An important fact to be noted here is that elliptic flow $\mathrm{v_{2}}$ arises not only from fluctuations in the initial stages of a collision but also from the initial-state geometry of the nuclear overlap region. In central collisions, fluctuations become quite important while the geometry plays its dominant role in peripheral collisions. Elliptic flow fluctuations due to the smaller size and non-spherical nuclear shape give quantitative increment which evidently can be seen from our results of isotopic collisions.

\begin{figure}[htbp]
\includegraphics[width=8.0cm,height=6.0cm]{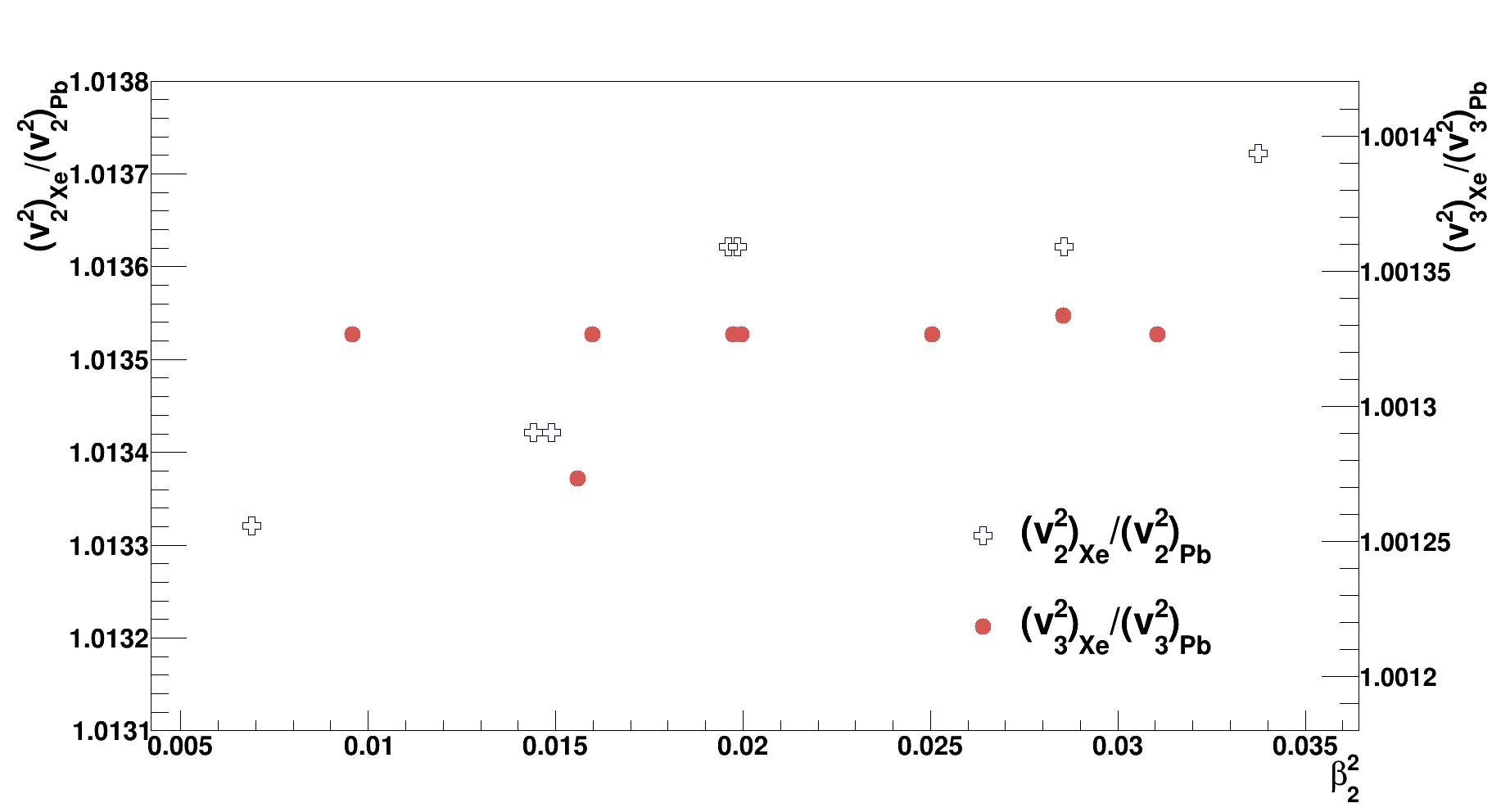}
\caption{Predicted ratio of $\mathrm{(v_{2}^{2})_{\mathrm{Xe}}/(v_{2}^{2})_{\mathrm{Pb}}}$ and $\mathrm{(v_{3}^{2})_{\mathrm{Xe}}/(v_{3}^{2})_{\mathrm{Pb}}}$ as a function of $\mathrm{(\beta_{2}^{2})_{\mathrm{Xe}}}$ in most-central collisions using Monte Carlo HYDJET++ model. The figure shows results for isotopes (A=128-135) of Xenon with different values of quadrupole moment.}
\label{Xesq_ratio_Pbsq}
\end{figure}

Figure \ref{Xesq_ratio_Pbsq} presents the predicted ratios $\mathrm{(v_{2}^{2})_{\mathrm{Xe}}/(v_{2}^{2})_{\mathrm{Pb}}}$ and $\mathrm{(v_{3}^{2})_{\mathrm{Xe}}/(v_{3}^{2})_{\mathrm{Pb}}}$ as a function of $\mathrm{(\beta_{2}^{2})_{\mathrm{Xe}}}$ where we find a strict inter-dependence. The ratio for elliptic flow varies linearly with $\mathrm{\beta_{2}^{2}}$ whereas for triangular flow, the ratio appears to be quite independent of $\mathrm{\beta_{2}^{2}}$, revolving around a constant value as quadrupole moment increases and atomic mass decreases.

\begin{figure}[htbp]
\includegraphics[width=8.5cm,height=6.0cm, keepaspectratio]{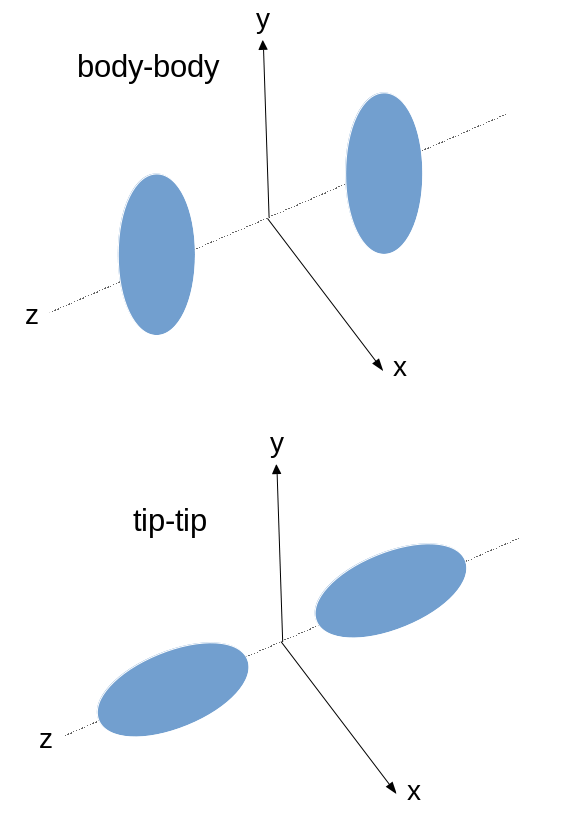} 
\caption{Collision scenario in body-body and tip-tip type geometrical configurations under the HYDJET++ framework \cite{Pandey_2022}.}
\label{configurations}
\end{figure} 

HYDJET++ model does not include nuclear density profiles for nuclei with some intrinsic deformation. Therefore, the nuclear density function needs to be modified for deformed nucleus like U, Xe, etc. In our present work, we have used the Modified Woods-Saxon (MWS) given by equation \ref{eq1} to calculate the initial distributions of partons, etc. for deformed nuclei \cite{PhysRevC.103.014903, Pandey_2022, refId0}. Within HYDJET++ framework,

$\mathrm{\rho_{0} =  \rho_{0}^{const} + \text{correction}}$,\\

$\mathrm{\rho_{0}^{const} = \frac{M}{V} =  \frac{3A}{4 \pi R'^{3}}}$,\\

$\text{The correction term is calculated as=}\mathrm{\rho_{0}^{const} × (\pi f/R')^{2}}$,\\ 

$\text{where } f= 0.54  \mathrm{fm} $,\\

$\mathrm{\beta_{2}}=\text{as per table \ref{table} , and}$\\ 

$\mathrm{\beta_{4}}=0.0 \text{ are the deformation parameters}$,\\

$\text{a = diffuseness parameter}= 0.59 \mathrm{fm}$,\\

$\mathrm{Y_{20} = \sqrt{\frac{5}{16\pi}}(3\cos ^{2}\theta-1)}, \text{and}$\\

$\mathrm{Y_{40} = \frac{3}{16\sqrt{\pi}}(35\cos ^{4}\theta-30\cos ^{2}\theta + 3)}$.\\

The values of different parameters have been taken from the reference \cite{MOLLER20161}. However, HYDJET++ model works in cylindrical coordinate system unlike the AMPT model that works in spherical coordinate sysytem. Thus, the modified nuclear density profile function in equation \ref{eq1} is converted from spherical polar coordinates ($\mathrm{r, \theta, \phi}$) to cylindrical polar coordinates ($\mathrm{\rho, z, \psi}$) using the relations $\mathrm{\theta =\tan^{-1} (z/r)}$ and $\mathrm{\theta=\tan^{-1} (r/z)}$ for body-body and tip-tip geometrical configurations, respectively. Here, $\mathrm{r}$ in the relations is basically $\mathrm{\rho}$ of cylindrical polar coordinate system and not spherical polar coordinate $\mathrm{r}$. This representation helps in eliminating the confusion of $\mathrm{\rho}$ in cylindrical polar coordinate system with the $\mathrm{\rho}$ in the nuclear density function. The Woods Saxon nuclear density profile in tip-type and body-type geometrical configurations is defined in the way two nuclei collide when approaching towards each other along the direction of propagation as shown in figure \ref{configurations}.

\begin{figure}[htbp]
\includegraphics[width=8.5cm,height=6.0cm, keepaspectratio]{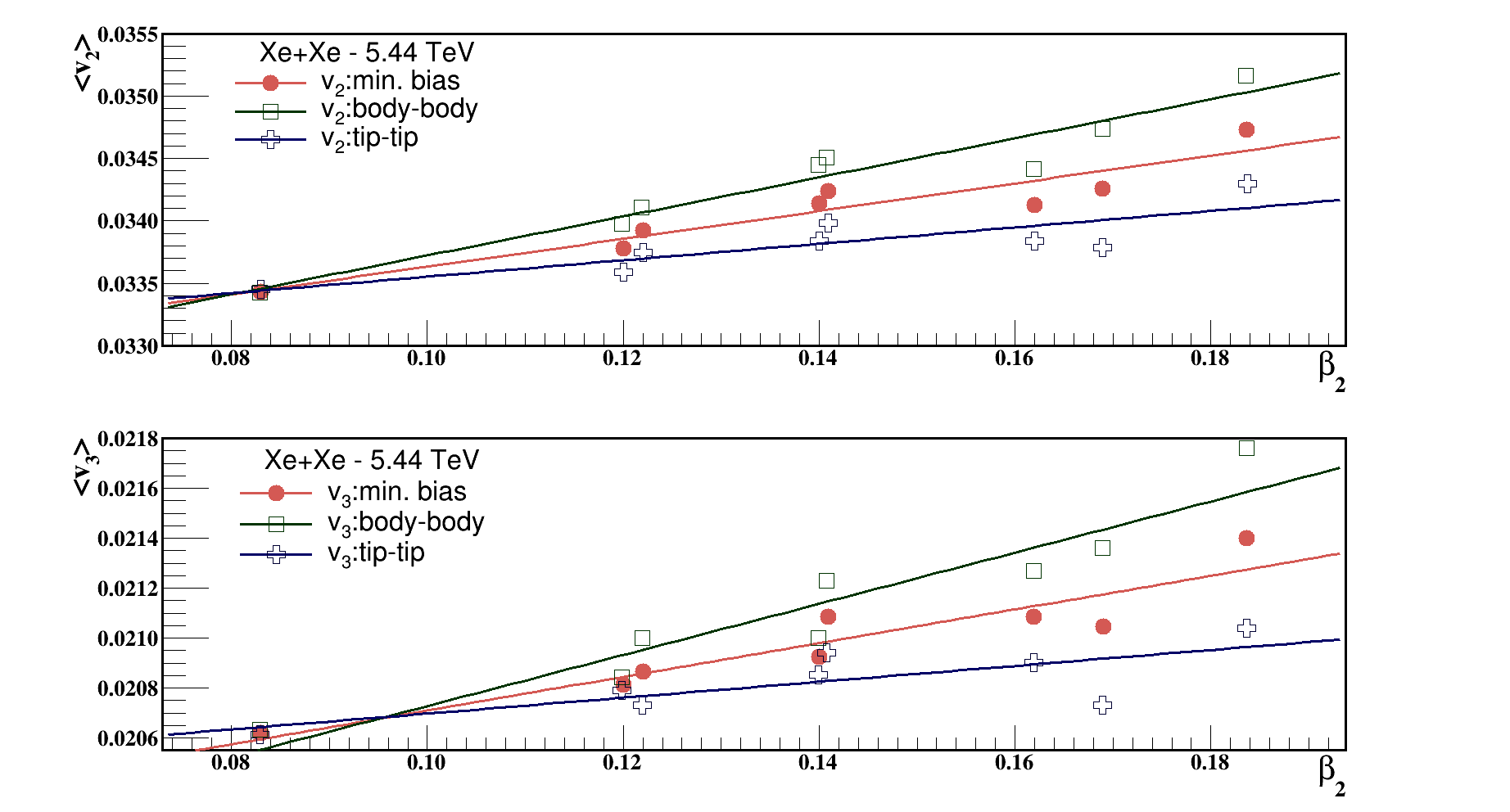} 
\caption{Variation of $\mathrm{p_{T}}$-integrated eliptic and triangular flow with respect to nuclear quadrupole moment $\mathrm{\beta_{2}}$ in deformed Xe-Xe collisions at 5.44 TeV center of mass energy using HYDJET++ model. The results have been presented in body-body and tip-tip type of geometric nucleus-nucleus collisions along with the minimum-bias results.}
\label{avgvn_beta2}
\end{figure}

In figure \ref{avgvn_beta2}, we present the $\mathrm{p_{T}}$-integrated anisotropic flow in deformed Xe-Xe collisions for the various isotopes at 5.44 TeV centre-of-mass energy. It is observed that, flow for body-body collisions is larger than flow for tip-tip collisions. Minimum bias collisions lie almost in mid-between these two extreme geometries. As nuclear quadrupole moment $\mathrm{\beta_{2}}$ increases or atomic mass decreases, flow increases. As a result, a linear inter-dependence is observed between the observables. The intersection point of the various types of collisions suggests that the geometrical configurations overlap or have same flow. At this point, it is difficult to disentangle the collisions on the basis of nuclear orientation just before collision. The intersection point of the various types of collisions is observed to be at a lower value of $\mathrm{\beta_{2}}$ for elliptic flow (intercept around 0.0333-0.0334) than for triangular flow (intercept around 0.0204-0.0206).

\begin{figure}[htbp]
\includegraphics[width=8.5cm,height=7.5cm, keepaspectratio]{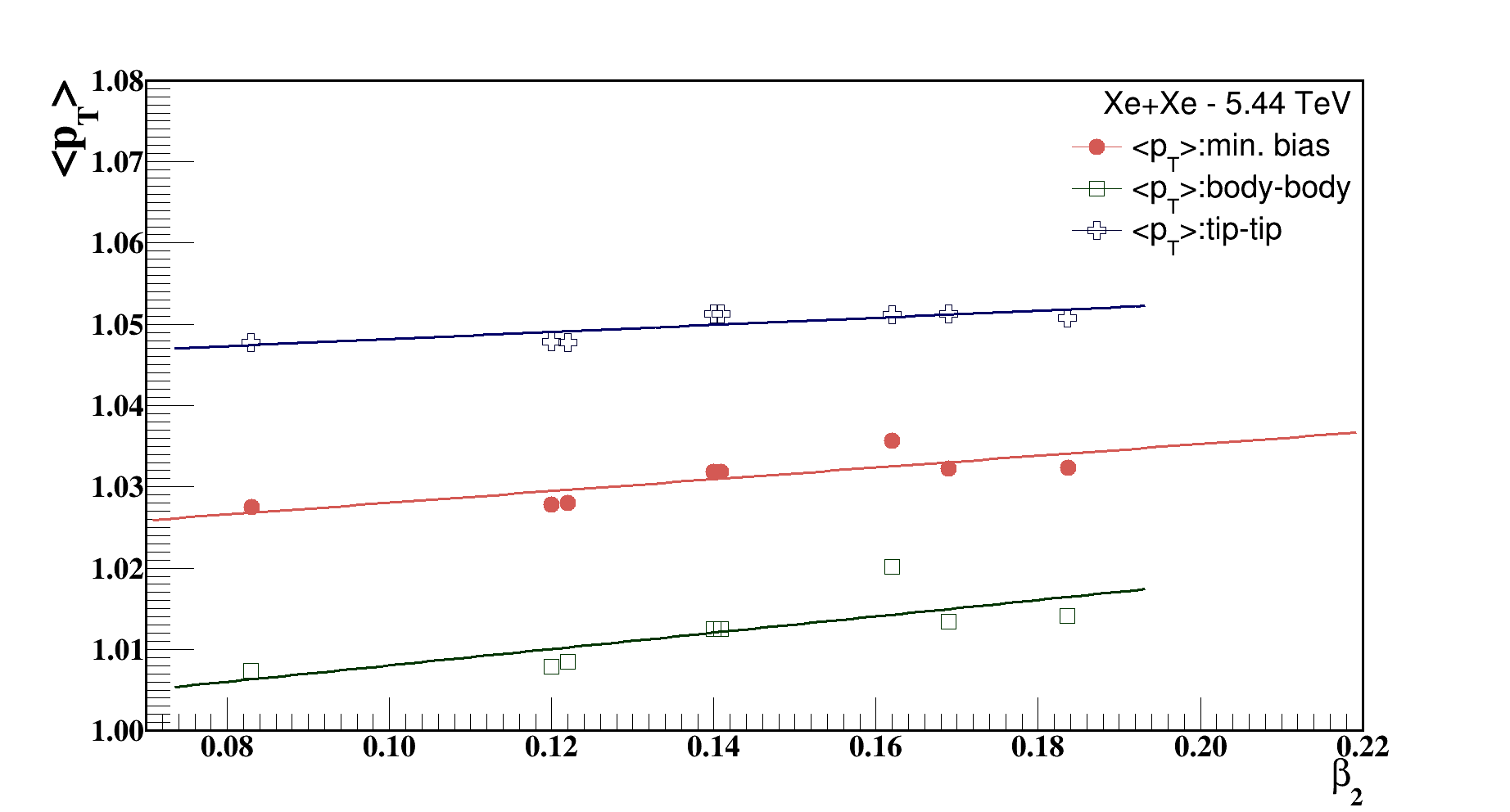}
\caption{Variation of mean transverse momentum $\mathrm{\langle p_{T} \rangle}$ as a function of nuclear quadrupole deformation parameter in deformed Xe-Xe collisions at 5.44 TeV center of mass energy using HYDJET++ model. The figure shows flow for isotopes (A=128-135) of Xenon with different values of quadrupole moment. The results have been presented in body-body and tip-tip type of geometric nucleus-nucleus collisions along with the minimum-bias results.}
\label{avgpt_beta2}
\end{figure}

\begin{figure}[htbp]
\includegraphics[width=8.5cm,height=6.0cm, keepaspectratio]{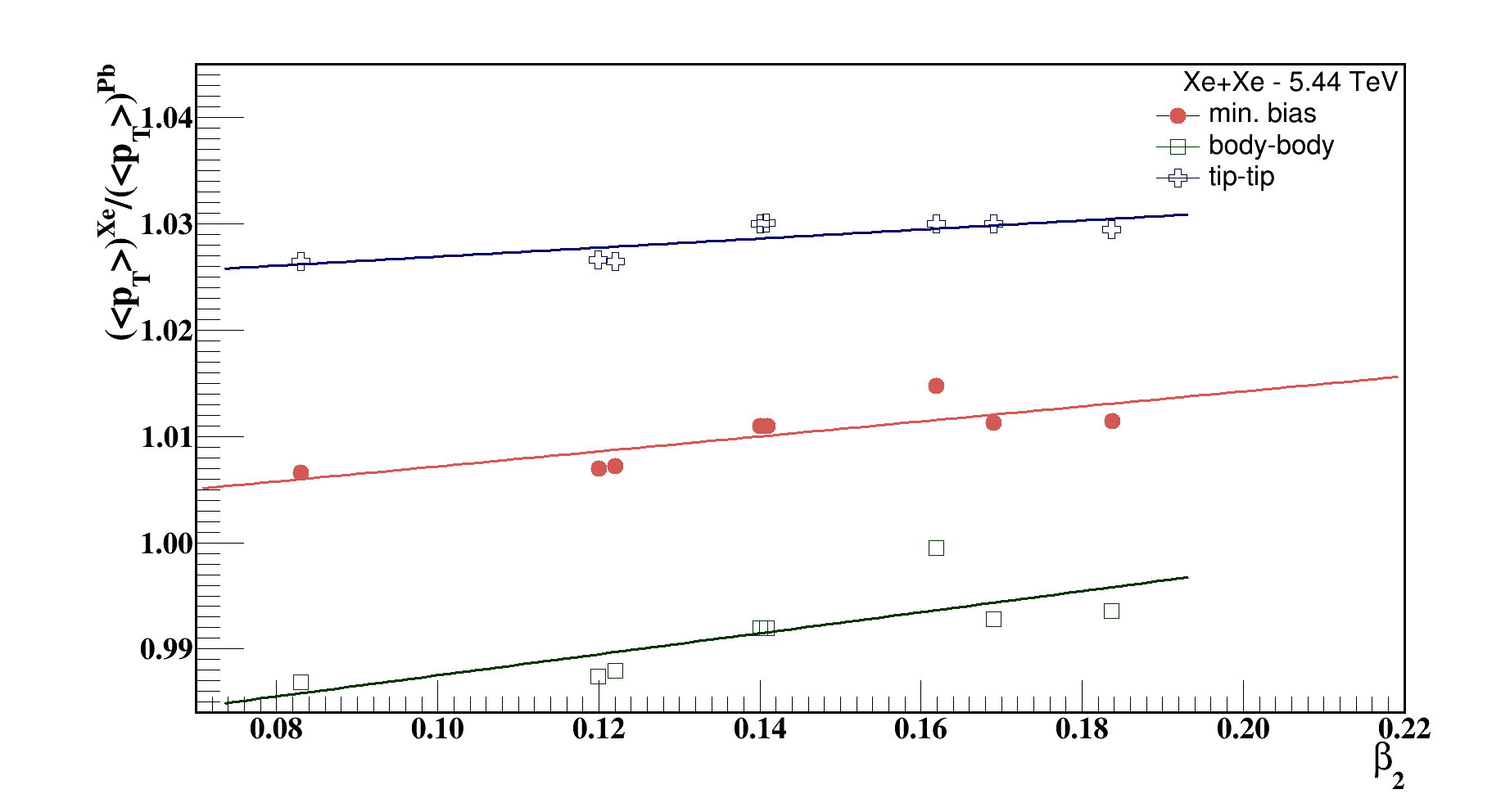} 
\caption{Scaling prediction of mean transverse momentum $\mathrm{\langle p_{T} \rangle}$ from deformed $\prescript{128-135}{54}{\mathrm{Xe}}$-$\prescript{128-135}{54}{\mathrm{Xe}}$ collisions to spherical $^{208}\mathrm{Pb}$-$^{208}\mathrm{Pb}$ collisions at 5.02 TeV center of mass energy using HYDJET++ model as a function of nuclear quadrupole deformation parameter. The figure shows results for isotopes (A=128-135) of Xenon with various values of nuclear deformation parameter ($\mathrm{\beta_{2}}$).}
\label{Xept_ratio_Pbpt}
\end{figure}

In order to study a proper shape-size correlation, we must also look into mean transverse momentum. $\mathrm{\langle p_{T} \rangle}$ carries essential information about the high-density deconfined state of strongly interacting matter, quark-gluon plasma. The slope of $\mathrm{p_{T}}$ distribution measures the inverse of the source temperature of the fireball, increases as one moves from most-central to most-peripheral collisions. This testifies the fact that the fireball temperature is maximum in most-central collisions \cite{PhysRevC.103.014903, refId0}. However, in deformed collisions it is quite exciting how does the deformation affects this fireball temperature. In figure \ref{avgpt_beta2}, we study the mean transverse momentum as a function of deformation parameter. Here, we observe that, with increasing deformation, mean transverse momentum increases, but not so strongly. In article \cite{GARDIM2020135749}, it is explained that in ultracentral collisions, the temperature increases as a function of the multiplicity, which in turn implies a rise of the mean transverse momentum of charged hadrons, $\mathrm{\langle p_{T} \rangle}$, observed in the final state, due to tight correlation with the temperature. As a result, the source temperature increases as nuclear deformation increases. Viewing from the door of geometry, we see that $\mathrm{\langle p_{T} \rangle}$ is higher for tip-tip collisions than body-body collisions. This implies that fireball temperature is higher in tip-tip collisions than in body-body collisions. Expressing in terms of system-size, we can say that, as system-size (atomic mass A) increases, mean $\mathrm{\langle p_{T} \rangle}$ decreases. This is more evidently visualized in figure \ref{Xept_ratio_Pbpt} where the ratio of mean transverse momentum in (deformed-) Xe-Xe collisions at 5.44 TeV to mean transverse momentum in spherical Pb-Pb collisions at 5.02 TeV is presented. It is observed that the ratio increases with increasing deformation and decreasing system-size. The mean transverse momentum in Xe-Xe collisions decreases by (1-2)\% compared to larger spherical Pb-Pb collisions at 5.02 TeV. Another important point to be noted is that the increasing $\mathrm{ \langle p_{T} \rangle}$ with nuclear deformation is attributed to the increasing transverse radial flow with increasing deformation; and as system-size in terms of atomic mass increases, this transverse radial flow decreases.

\section{Summary and Outlook}
\label{summary}

In summary, we performed a study on some isotopes of xenon using Monte-carlo technique under HYDJET++ framework. Here, we studied two observables, anisotropic flow and transverse momentum, as a function of the two important properties of an isotope, namely the nuclear (quadrupole) deformation and system-size as in atomic mass, ($\mathrm{A}$). Anisotropic flow for odd-A isotopes is suppressed than flow for even-A isotopes. As a function of transverse momentum, for even isotopes, anisotropic flow is directly proportional to nuclear deformation whereas for odd isotopes, this flow is inversely proportional to nuclear quadrupole moment. The ratios of elliptic flow and triangular flow  of deformed collisions with respect to their respective spherical collisions present a linear relationship with increasing transverse momentum and gradually decreasing towards higher $\mathrm{p_{T}}$ region. Large quadrupole moment gives rise to higher flow for even-A isotopic collisions, hence, higher ratios. Opposite happens for odd-A isotopic collisions. The effect of nuclear deformation is visible maximum in low $\mathrm{p_{T}}$ region and decreases moving towards higher $\mathrm{p_{T}}$ values. Xenon, being deformed gave us an opportunity to present our study in two extreme geometries of collisions, body-body and tip-tip type of Xe-Xe collisions at 5.44 TeV. We found a linear inter-dependence between $\mathrm{p_{T}}$ integrated anisotropic flow and nuclear deformation. Mean transverse momentum marks the fireball temperature in body-body and tip-tip collisions. $\mathrm{\langle p_{T} \rangle}$ is higher in tip-tip collisions than body-body collisions. There exists a negative linear correlation of $\mathrm{\langle p_{T} \rangle}$ with collision system-size appears and a positive correlation with nuclear deformation. \\

The modeling of heavy-ion collisions helps in pulling up the hydrodynamic response of $\mathrm{v_{n}}$ to the contribution of nuclear deformation and thereby to the eccentricity, which is quite different from the case of spherical nuclei collision systems. Thus, we see that heavy-ion isotopic collisions offer immense potential to constrain the nuclear deformation parameters and help in testing the state-of-the-art nuclear structure models. Our measurements of $v_{2}$ and $\mathrm{v_{3}}$ show strong differences between deformed isotopic $\prescript{128-135}{54}{\mathrm{Xe}}$-$\prescript{128-135}{54}{\mathrm{Xe}}$ collisions and spherical $^{208}\mathrm{Pb}$-$^{208}\mathrm{Pb}$ collisions. These differences are well explained from the quadrupole deformation $\mathrm{\beta_{2}}$ in $\prescript{128-135}{54}{\mathrm{Xe}}$ isotopes. However, this deformation parameter can be a more useful tool to constrain the initial-state geometry as well as the shape of the nucleus, for example the study of initial state fluctuations in anisotropic flow might curb the geometry of xenon nucleus being either prolate or oblate or purely triaxial such as in \cite{PhysRevLett.128.082301}. This we preserve for future work. Hence, isotopic heavy-ion collisions act as a precision tool to describe the role of nuclear quadrupole deformation in the field of heavy-ion physics and will provide strong support to its role which is quite debative in nuclear structure physics. Such analysis may be hopefully done at the relativistic high-energy collider experiments at LHC in future but for that individual study of either isotope in the field heavy-ion collisions is necessary as a background study. 

\section{ACKNOWLEDGEMENTS}
BKS sincerely acknowledges financial support from the Institutions of Eminence (IoE) BHU grant number-6031. Saraswati Pandey acknowledges the financial support obtained from UGC and IoE under a research fellowship scheme during the work.

\end{document}